\documentclass[aps,pra,twocolumn,groupedaddress,nofootinbib]{revtex4-1}
\usepackage{amsmath}
\usepackage{amssymb}
\usepackage{graphicx,subfigure}
\usepackage{mathrsfs}
\usepackage{ntheorem}
\usepackage[T1]{fontenc}
\usepackage{bm}
\usepackage{subfigure}
\usepackage[bookmarks=false]{hyperref}
\usepackage{xcolor}
\hypersetup{colorlinks=true,citecolor=blue,
linkcolor=blue,urlcolor=blue,pdfstartview=FitH,
bookmarksopen=true}

\newcommand{\ket}[1]{|#1\rangle}
\newcommand{\bra}[1]{\langle #1|}
\newcommand{\inp}[2]{\langle #1|#2\rangle}
\newcommand{\inpp}[2]{\prec #1,#2 \succ}

\newcommand{\expt}[1]{\langle #1 \rangle}

\newcommand{\tr}{\mathrm{tr}}
\newcommand{\ud}{\mathrm{d}}

\newcommand{\abs}[1]{\lvert #1\rvert}

\def\CC{{\rm\kern.24em \vrule width.04em height1.46ex depth-.07ex \kern-.30em C}}
\def\RR{{\rm\kern.24em \vrule width.04em height1.46ex depth-.07ex
\kern-.30em R}}
\def\P{{\rm I\kern-.25em P}}

\begin{document}

\title{Time-dependent $\mathcal{PT}$-symmetric quantum mechanics in generic non-Hermitian systems}

\author{Da-Jian Zhang}
\affiliation{Department of Physics, National University of Singapore, Singapore 117542}
\author{Qing-hai Wang}
\affiliation{Department of Physics, National University of Singapore, Singapore 117542}
\author{Jiangbin Gong}
\email{phygj@nus.edu.sg}
\affiliation{Department of Physics, National University of Singapore, Singapore 117542}
\date{\today}

\begin{abstract}
$\mathcal{PT}$-symmetric quantum mechanics has been considered an important theoretical framework for understanding physical phenomena in $\mathcal{PT}$-symmetric systems, with a number of $\mathcal{PT}$-symmetry related applications.  This line of research was made possible by the introduction of a time-independent metric operator to redefine the inner product of a Hilbert space.
To treat the dynamics of generic non-Hermitian systems under equal footing, we advocate in this work the use of
a time-dependent metric operator for the inner-product between time-evolving states.  This treatment makes it possible to always interpret the dynamics of arbitrary (finite-dimensional) non-Hermitian systems in the framework of time-dependent $\mathcal{PT}$-symmetric quantum mechanics, with unitary time evolution, real eigenvalues of an energy observable, and quantum measurement postulate all restored.   Our work sheds new lights on generic non-Hermitian systems and spontaneous $\mathcal{PT}$-symmetry breaking in particular.  We also illustrate possible applications of our formulation with well-known examples in quantum thermodynamics.
\end{abstract}

\maketitle

\section{Introduction}\label{sec:in}

Isolated quantum systems are described by Hermitian Hamiltonians, yet no quantum system is completely isolated in reality. Quite often the dynamics of an open quantum system permits an effective description based on a non-Hermitian Hamiltonian:
\begin{eqnarray}\label{SE}
i\partial_t\ket{\psi(t)}=\mathcal{H}(t)\ket{\psi(t)},
\end{eqnarray}
provided that some
appropriate conditions are met \cite{1992Dalibard580,2014Daley77}. (Units such that $\hbar=1$ are adopted in this paper.) Note that the effective Hamiltonian in such treatment is time-dependent in general. Equation (\ref{SE}) can be used to describe a broad spectrum of dissipative systems \cite{2005Razavy}, wave systems subject to gain and loss \cite{1998Ching1545,2015Cao61}, and solid-state systems with non-Hermitian self-energy stemming from electron-electron interactions
or disorders \cite{2017Kozii,2018Papaj,2018Shen26403}. With these physical motivations, studies of the dynamics of non-Hermitian systems as described by Eq.~(\ref{SE}) have been fruitful \cite{2011Moiseyev}.

Since the pioneering work of Bender and Boettcher \cite{1998Bender5243}, considerable research effort, both theoretical and experimental, has been devoted to a subclass of non-Hermitian systems that are described by time-independent effective Hamiltonians $\mathcal{H}_{\cal PT}$ respecting parity-time reversal ($\mathcal{PT}$) symmetry \cite{2016Konotop35002,2017Feng752,2018El-Ganainy11}. This is stimulated by the discovery that a quantum theory, known as $\mathcal{PT}$-symmetric quantum mechanics ($\mathcal{PT}$QM), can be built for these non-Hermitian systems \cite{2002Bender270401,2002Mostafazadeh205}. One main conceptual advance of $\mathcal{PT}$QM is the introduction of a time-independent metric operator $\mathcal{W}$ to redefine the inner product of a Hilbert space. $\mathcal{W}$ is chosen in such a way that
\begin{eqnarray}\label{Bender-WH}
\mathcal{W}\mathcal{H}_{\cal PT}=\mathcal{H}_{\cal PT}^\dagger \mathcal{W},
\end{eqnarray}
because of which $\mathcal{H}_{\cal PT}$ is rendered Hermitian under the new inner product. The theory is thus not in conflict with
standard quantum mechanics but rather becomes its complex generalization.
To date, $\mathcal{PT}$QM has been considered an important theoretical framework for understanding physical phenomena involving time-independent $\mathcal{PT}$-symmetric Hamiltonians, leading to fascinating applications, such as $\mathcal{PT}$-symmetric synthetic photonic lattices
\cite{2012Regensburger167},
non-reciprocal light propagation \cite{2013Feng108,2011Feng729,2014Peng394}, and single-mode lasers \cite{2014Feng972,2014Hodaei975}.  Recently $\mathcal{PT}$QM has been experimentally realized in a genuine quantum system \cite{Du2019}.

Given the importance of $\mathcal{PT}$QM, the next natural step is to extend it to a broader class of non-Hermitian systems. This issue is important but challenging, because of the no-go theorem \cite{2007Mostafazadeh208}, which claims that the unitarity of Eq.~(\ref{SE}) with $\mathcal{H}(t)$ being $\mathcal{PT}$-symmetric and the existence of time-dependent metric operator are incompatible in general. Many attempts have been made towards solving this issue \cite{2003Mostafazadeh155,2004Mostafazadeh1,2008Znojil85003,2013Gong485302,
2016Fring42114,2016Fring42128,2017Fring2318,2018Mostafazadeh46022}. A representative one among them is Gong and Wang's proposal \cite{2013Gong485302} for extending $\mathcal{PT}$QM to the time-dependent Hamiltonian $\mathcal{H}_{\cal PT}(t)$ that satisfies
\begin{eqnarray}\label{time-dependent-WH}
\mathcal{W}(t)\mathcal{H}_{\cal PT}(t)=\mathcal{H}_{\cal PT}^\dagger(t)\mathcal{W}(t),
\end{eqnarray}
for a certain time-dependent metric operator $\mathcal{W}(t)$. In overcoming the no-go theorem, Gong and Wang advocated a Schr\"{o}dinger-like equation
\begin{eqnarray}\label{SE-like}
i\partial_t\ket{\psi(t)}=\left[\mathcal{H}_{\cal PT}(t)-\frac{i}{2}\mathcal{W}^{-1}(t)
\partial_t\mathcal{W}(t)
\right]\ket{\psi(t)},
\end{eqnarray}
where an additional term $-\frac{i}{2}\mathcal{W}^{-1}(t)
\partial_t\mathcal{W}(t)$ emerges as compared with Eq.~(\ref{SE}) in order to restore the unitarity of time evolution. Equation (\ref{SE-like}) has proven to be useful and spurred reexaminations of interesting issues in quantum dynamics and thermodynamics \cite{2016Gardas23408,2017Maamache383,2017Mead85001,2018Mostafazadeh46022,
2018Wei12105,2018Wei12114}, including the well-known Jarzynski equality (JE) \cite{2015Deffner150601} and Crooks fluctuation theorem (CFT) \cite{2017Zeng31001}.

Despite the usefulness of Eq.~(\ref{SE-like}), the dynamics of a non-Hermitian system is often given in terms of Eq.~(\ref{SE}) rather than the form of Eq.~(\ref{SE-like}). Nevertheless, Ref.~\cite{2013Gong485302} indicates an enlightening viewpoint on understanding the role of the generator of a Schr\"{o}dinger equation. That is, given a Schr\"{o}dinger equation with a generator [e.g., $\mathcal{H}_{\cal PT}(t)-\frac{i}{2}\mathcal{W}^{-1}(t)
\partial_t\mathcal{W}(t)$ appearing in Eq.~(\ref{SE-like})],
this generator itself may not be naively thought of as the energy observable (Hamiltonian) operator associated with this Schr\"{o}dinger equation. Rather, only a part of it [e.g., $\mathcal{H}_{\cal PT}(t)$ appearing in Eq.~(\ref{SE-like})] may be interpreted as the Hamiltonian operator.
Inspired by this viewpoint, we here advocate an alternative approach to overcoming the no-go theorem. Instead of adding a new term in Eq.~(\ref{SE}), we decompose $\mathcal{H}(t)$ into two parts. To be specific, we find a suitable time-dependent metric operator, denoted as $W(t)$, and decompose  $\mathcal{H}(t)$ as
\begin{eqnarray}\label{decomposition}
\mathcal{H}(t)=H(t)+iK(t),
\end{eqnarray}
where $H(t)$ satisfies
\begin{eqnarray}\label{our-WH}
W(t)H(t)=H^\dagger(t)W(t),
\end{eqnarray}
and $K(t)=-\frac{1}{2}W^{-1}(t)\partial_tW(t)$. Clearly, $H(t)$ and $iK(t)$ correspond to $\mathcal{H}_{\cal PT}(t)$ and $-\frac{i}{2}\mathcal{W}^{-1}(t)
\partial_t\mathcal{W}(t)$ appearing in Eq.~(\ref{SE-like}), respectively. Analogous to that in Ref.~\cite{2013Gong485302}, the idea here is that we no longer treat the generator of Eq.~(\ref{SE}) itself, i.e., $\mathcal{H}(t)$, as the genuine Hamiltonian operator (or, in other words, the energy observable). Instead, we revise the energy observable to be only a part of it, i.e., $H(t)$, since $H(t)$ satisfies Eq.~(\ref{our-WH}) whereas $\mathcal{H}(t)$ does not. This revision enables us to restore the unitarity of time evolution, the reality of energy eigenvalues, and the quantum measurement postulate, while retaining Eq.~(\ref{SE}) as the Schr\"{o}dinger equation of a non-Hermitian system.

Our result is significant, since it is applicable to arbitrary (finite-dimensional) non-Hermitian systems. To our best knowledge, the possibility that $\mathcal{PT}$QM can be extended to generic (finite-dimensional) non-Hermitian systems has never been realized or even imagined before. Indeed, this possibility seemed to be ruled out by  previous studies on $\mathcal{PT}$-symmetry breaking and thermodynamical relations such as JE \cite{2015Deffner150601} and CFT \cite{2017Zeng31001}, since they have found significant differences between non-Hermitian systems with and without $\mathcal{PT}$ symmetry when studying quantum thermodynamics. Because of this,  we shall revisit these studies and show that JE and CFT can be extended to arbitrary (finite-dimensional) non-Hermitian systems.

This paper is organised as follows. In Sec.~\ref{sec:tf}, we provides our theoretical framework of extending $\mathcal{PT}$QM to generic non-Hermitian systems. In Sec.~\ref{sec:de}, we discuss an alternative view of $\mathcal{PT}$-symmetry breaking. In Sec.~\ref{sec:im}, we presents two applications in quantum thermodynamics. We conclude this work in Sec.~\ref{sec:con} with some remarks.

\section{theoretical framework}
\label{sec:tf}

\subsection{Time-independent $\mathcal{PT}$-symmetric systems}
To present our theory clearly and self-consistently, we first recapitulate some fundamentals of $\mathcal{PT}$QM for time-independent problems. Consider a quantum system $S$ with a time-independent $\mathcal{PT}$-symmetric Hamiltonian $\mathcal{H}$. In the following, $\mathfrak{H}_S$ [$d:=\textrm{dim}(\mathfrak{H}_S)<\infty$] is used to denote the Hilbert space of $S$, and $\ket{\varphi_i}$ and $\bra{\phi_i}$ the eigenket and eigenbra of $\mathcal{H}$, i.e.,
\begin{eqnarray}
\mathcal{H}\ket{\varphi_i}=\omega_i\ket{\varphi_i},~~
\bra{\phi_i}\mathcal{H}=\omega_i\bra{\phi_i},
\end{eqnarray}
where $\omega_i\in\mathbb{C}$ is the associated eigenvalue. Contrary to the Hermitian case, the $\ket{\varphi_i}$'s are not orthogonal to each other, nor are the $\bra{\phi_i}$'s, and furthermore $\ket{\varphi_i}\neq\ket{\phi_i}$ in general. A remarkable observation made by Bender \textit{et al.}~\cite{2002Bender270401} is that in the case of unbroken $\mathcal{PT}$ symmetry, the $\ket{\varphi_i}$'s and $\bra{\phi_i}$'s each form a complete set, and moreover, all the $\omega_i$'s are real. One is thus led to the introduction of a positive-definite metric operator
\begin{eqnarray}\label{Bender-MO}
\mathcal{W}:=\sum_{i=1}^d\ket{\phi_i}\bra{\phi_i}
\end{eqnarray}
to redefine the inner product of $\mathfrak{H}_S$ as
\begin{eqnarray}\label{Bender-IP}
\inpp{\cdot}{\cdot}:=\bra{\cdot}\mathcal{W}\ket{\cdot}.
\end{eqnarray}
Initially the metric operator is connected with the parity operator $\mathcal{P}$, the time-reversal operator $\mathcal{T}$, and the so-called $\mathcal{C}$ operator \cite{2002Bender270401}. Here we take it as an abstract metric.
Under the new inner product (\ref{Bender-IP}), $\mathcal{H}$ is rendered Hermitian, $\inpp{\cdot}{\mathcal{H}\cdot}=\inpp{\mathcal{H}
\cdot}{\cdot}$ or equivalently $\mathcal{W}\mathcal{H}=
\mathcal{H}^\dagger \mathcal{W}$ [i.e., Eq.~(\ref{Bender-WH})]. Then the unitarity of time evolution is restored because of the Hermiticity of $\mathcal{H}$; that is, $\inpp{\psi(t)}{\psi(t)}=\textrm{constant}$, with $\ket{\psi(t)}$ denoting an arbitrary evolving state of $S$.
A quantum theory, namely, $\mathcal{PT}$QM, is built consistently by identifying any observable with a Hermitian operator thus defined. In passing, $\mathcal{W}$ is dynamically determined as it implicitly depends on the Hamiltonian $\mathcal{H}$, and moreover, its choices are not unique.

\subsection{Generic non-Hermitian systems}

Suppose now that $S$ is described by a generic effective Hamiltonian $\mathcal{H}(t)$, typically depending on the time $t$. For this more general situation, motivated by Bender \textit{et al.}'s  pioneering work, we aim to establish a quantum theory while retaining the unitarity of time evolution, the reality of energy eigenvalues, and an extended quantum measurement postulate.
Our idea is to construct a suitable \textit{time-dependent} metric operator, $W(t)$, from a given $\mathcal{H}(t)$.

\subsubsection{The time-dependent metric operator}

To restore the unitarity of time evolution (\ref{SE}), we require that $W(t)$ is positive-definite and satisfies $\bra{\psi(t)}W(t)\ket{\psi(t)}=\textrm{constant}$ for any evolving state.  If we redefine the inner product of $\mathfrak{H}_S$ through use of $W(t)$,
\begin{eqnarray}\label{inner-product}
\inpp{\cdot}{\cdot}_t:=\bra{\cdot}W(t)\ket{\cdot},
\end{eqnarray}
then this requirement amounts to the conservation of the inner product of $\ket{\psi(t)}$ with itself.
The subscript $t$ appearing in Eq.~(\ref{inner-product}) is used to indicate the time-dependence of this inner product.
In order to meet our requirement, it is sufficient and necessary for $W(t)$ to be of the following form
\begin{eqnarray}\label{metric-operator}
W(t)=\eta^\dagger(t)\eta(t),
\end{eqnarray}
with $\eta(t)$ satisfying
\begin{eqnarray}\label{eta}
i\frac{d}{d t}\eta(t)=-\eta(t)\mathcal{H}(t).
\end{eqnarray}
Indeed, as a solution to Eq.~(\ref{eta}), $\eta(t)$ must be invertible. In conjunction with Eq.~(\ref{metric-operator}), this implies that $W(t)$ is positive-definite. Besides,
using Eqs.~(\ref{SE}) and (\ref{eta}), we have $\partial_t[\eta(t)\ket{\psi(t)}]=0$. It follows that $\eta(t)\ket{\psi(t)}=\textrm{constant}$, which further leads to  $\bra{\psi(t)}W(t)\ket{\psi(t)}=\textrm{constant}$. This proves the sufficiency. To prove the necessity, we deduce from the positive-definiteness of $W(t)$ that there exists an invertible operator $\widetilde{\eta}(t)$ such that $W(t)=\widetilde{\eta}^\dagger(t)\widetilde{\eta}(t)$. Letting $\ket{\widetilde{\psi}(t)}:=\widetilde{\eta}(t)\ket{\psi(t)}$ and using $\bra{\psi(t)}W(t)\ket{\psi(t)}=\textrm{constant}$, we have $\inp{\widetilde{\psi}(t)}{\widetilde{\psi}(t)}=\textrm{constant}$. This indicates that $\ket{\widetilde{\psi}(t)}$ undergoes a unitary evolution. Denote the corresponding evolution operator as $U(t)$, i.e., $\ket{\widetilde{\psi}(t)}=U(t)\ket{\widetilde{\psi}(0)}$, and define $\eta(t):=U^\dagger(t)\widetilde{\eta}(t)$. Simple calculations show that $\eta(t)\ket{\psi(t)}=\ket{\widetilde{\psi}(0)}$. Differentiating both sides of this equality and using Eq.~(\ref{SE}), we have that $\eta(t)$ fulfills Eq.~(\ref{eta}). So, there is
\begin{eqnarray}
W(t)=\widetilde{\eta}^\dagger(t)\widetilde{\eta}(t)=\eta^\dagger(t)\eta(t),
\end{eqnarray}
with $\eta(t)$ satisfying Eq.~(\ref{eta}). This completes the proof.

A few remarks are in order.
First,
just like how the metric operator in Eq.~(\ref{Bender-MO}) is defined, our metric operator is  dynamically
determined, as can be seen from Eqs.~(\ref{metric-operator}) and (\ref{eta}). Second, the choices of $W(t)$ are not unique, but the applications of our framework may be irrespective of the specific choice adopted. Third, a time-independent effective Hamiltonian $\mathcal{H}$ does not necessarily produce a time-independent metric operator. Forth, a periodic $\mathcal{H}(t)$ [i.e., $\mathcal{H}(t+\tau)=\mathcal{H}(t)$] may or may not produce a periodic $W(t)$ with the same periodicity.

 \subsubsection{The energy observable}
To restore the reality of energy eigenvalues, we attempt to decompose  $\mathcal{H}(t)$ into two parts,
$\mathcal{H}(t)=H(t)+iK(t)$ [i.e., Eq.~(\ref{decomposition})],
where $H(t)$ and $K(t)$ are required to be physically Hermitian. Here and henceforth, an operator $X(t)$ is said to be physically Hermitian if it satisfies
\begin{eqnarray}
\inpp{\cdot}{X(t)\cdot}_t=\inpp{X(t)\cdot}{\cdot}_t
\end{eqnarray}
or equivalently
\begin{eqnarray}\label{pseudo-H}
W(t)X(t)=X^\dagger(t)W(t).
\end{eqnarray}
To figure out the explicit expressions of $H(t)$ and $K(t)$, we differentiate both sides of the equality $\bra{\psi(t)}W(t)\ket{\psi(t)}=\textrm{constant}$, and obtain
\begin{eqnarray}\label{step2}
\partial_tW(t)+i\left[\mathcal{H}^\dagger(t)W(t)
-W(t)\mathcal{H}(t)\right]=0.
\end{eqnarray}
Substituting Eq.~(\ref{decomposition}) into Eq.~(\ref{step2}) and using the Hermiticity of $H(t)$ and $K(t)$ under $W(t)$, we recover, after some simple algebra, the result $K(t)=-\frac{1}{2}W^{-1}(t)\partial_tW(t)$, as claimed in Sec.~\ref{sec:in}.

Further, it follows from Eq.~(\ref{decomposition}) that
\begin{eqnarray}\label{H(t)}
H(t)=\mathcal{H}(t)+\frac{i}{2}W^{-1}(t)\partial_tW(t).
\end{eqnarray}

Note that previous studies \cite{2013Gong485302,2018Mostafazadeh46022} have shown that for a time-varying inner product metric, the generator of a Schr\"{o}dinger equation should be comprised of a geometric part, which is determined by a metric-compatible connection on an underlying Hermitian vector bundle, and a
non-geometric part, which can be identified with the energy observable. Our derivations above make such insight much more general, because our starting point is a generic non-Hermitian $\mathcal{H}(t)$, with $W(t)$ not given a priori.
In light of these studies and noting that the obtained $K(t)$ is essentially the geometric part \cite{2013Gong485302,2018Mostafazadeh46022}, we regard $H(t)$ as the energy observable of our system $S$. Consequently, the energy eigenvalues are the eigenvalues of $H(t)$, which are always real because $H(t)$ is indeed Hermitian under the metric $W(t)$.

\subsubsection{The measurement postulate}

To complete our theory, we now address the measurement problem. Analogous to that in $\mathcal{PT}$QM, an observable in our theory, represented by the symbol $A_S(t)$, is identified with a physically Hermitian operator. This is consistent with our foregoing treatment of $H(t)$ as the energy observable.
Without loss of generality, $A_S(t)$ has a spectral decomposition
\begin{eqnarray}\label{AS}
A_S(t)=\sum_{i=1}^da_i(t)\ket{a_i(t)}\bra{a_i(t)}W(t).
\end{eqnarray}
Here, $a_i(t)\in\mathbb{R}$ denotes the eigenvalue and $\ket{a_i(t)}$ represents the associated eigenvector, satisfying
\begin{eqnarray}
\inpp{a_i(t)}{a_j(t)}_t=\delta_{ij},
\end{eqnarray}
where $\delta_{ij}$ denotes the Kronecker $\delta$ symbol.

A density operator in our theory, denoted as $\rho_S(t)$, is identified with a physically Hermitian operator that is positive-semidefinite w.r.t.~the inner product in Eq.~(\ref{inner-product}) and satisfies $\tr[\rho_S(t)]=1$. Accordingly, the density operator associated with the evolving state $\ket{\psi(t)}$ may be defined as
\begin{eqnarray}
\rho_S(t)=\ket{\psi(t)}\bra{\psi(t)}W(t),
\end{eqnarray}
where the normalization condition $\inpp{\psi(t)}{\psi(t)}_t=1$ has been assumed.
More generally, if $S$ is initially prepared in an ensemble of states $\{p_i,\ket{\psi_i(0)},i=1,\cdots,n\}$, with $\inpp{\psi_i(0)}{\psi_i(0)}_0=1$ and $p_i>0$ satisfying $\sum_{i=1}^{n} p_i=1$, the corresponding density operator is
\begin{eqnarray}\label{rho-ensemble}
\rho_S(t)=\sum_{i=1}^{n} p_i\ket{\psi_i(t)}\bra{\psi_i(t)}W(t),
\end{eqnarray}
where $\ket{\psi_i(t)}$ denotes the evolving state starting from $\ket{\psi_i(0)}$.
Direct calculations show that $\rho_S(t)$ satisfies
\begin{eqnarray}\label{LE}
i\partial_t\rho_S(t)=[\mathcal{H}(t),\rho_S(t)],
\end{eqnarray}
which may be regarded as the Liouville-von Neumann equation corresponding to Eq.~(\ref{SE}).

Now, a measurement in our theory is associated with an observable $A_S(t)$ and a density operator $\rho_S(t)$. We postulate that the possible outcomes of the measurement are the eigenvalues $a_i(t)$. Upon measuring $\rho_S(t)$, the probability of getting the outcome $a_i(t)$ is postulated to be
\begin{eqnarray}
P_i(t)=\tr\left[\ket{a_i(t)}\bra{a_i(t)}W(t)\rho_S(t)\right].
\end{eqnarray}
Accordingly,
the expectation value of $A_S(t)$ in the state $\rho_S(t)$ reads
\begin{eqnarray}
\sum_{i=1}^d a_i(t)P_i(t)=\tr[A_S(t)\rho_S(t)].
\end{eqnarray}
It is easy to see that this postulate meets the natural requirement that the expectation value of an observable must be real.

To gain further physical insight, we dilate and reinterpret $S$ as a part of a Hermitian system and show that the measurement postulated above can be realized within the standard quantum measurement theory. That is, corresponding to every measurement postulated above, there is a standard quantum measurement on the Hermitian system with the same outcomes $a_i(t)$ and the probabilities $P_i(t)$.

Specifically, the Hermitian system lives in the Hilbert space $\mathfrak{H}_A\otimes\mathfrak{H}_S$, with $\mathfrak{H}_A\simeq\mathbb{C}^2$ being an auxiliary Hilbert space. From now on, let us consider the evolution of $S$ during a finite time interval $[0,\tau]$. For this, there always exists a $W(t)$ such that $W(t)\geq I$ for all $t\in[0,\tau]$ \cite{1note}, where $I$ denotes the identity operator. So, the operator
\begin{eqnarray}\label{M}
M(t):=\sqrt{W(t)-I},
\end{eqnarray}
is well-defined, and moreover, it is Hermitian in the conventional sense.
Let $\ket{\widetilde{\psi}_i(t)}\in\mathfrak{H}_S$, $i=1,\cdots,d$, be any states satisfying Eq.~(\ref{SE}) and $\inpp{\widetilde{\psi}_i(t)}{\widetilde{\psi}_j(t)}_t=\delta_{ij}$. Using Eq.~(\ref{M}), one can verify that the following $2d$ states
\begin{eqnarray}
\ket{\widetilde{\Psi}_i(t)}:=\ket{0}\otimes\ket{\widetilde{\psi}_i(t)}+
\ket{1}\otimes
M(t)\ket{\widetilde{\psi}_i(t)},\label{MP:auxiliary-states}\\
\ket{\widetilde{\Psi}_{i+d}(t)}:=\ket{1}\otimes\ket{\widetilde{\psi}_i(t)}-
\ket{0}\otimes
M(t)\ket{\widetilde{\psi}_i(t)},
\end{eqnarray}
$i=1,\cdots,d$, form an orthonormal basis (w.r.t.~the conventional inner product in standard quantum mechanics) for $\mathfrak{H}_A\otimes\mathfrak{H}_S$. It follows that
\begin{eqnarray}
\widetilde{H}(t):=i\sum_{i=1}^{2d}\ket{\partial_t\widetilde{\Psi}_i(t)}
\bra{\widetilde{\Psi}_i(t)}
\end{eqnarray}
is a Hermitian operator in the conventional sense, which we regard as the Hamiltonian of the Hermitian system. That is, the Schr\"{o}dinger equation of the Hermitian system reads
\begin{eqnarray}\label{SE-AS}
i\partial_t\ket{\Psi(t)}=\widetilde{H}(t)\ket{\Psi(t)}.
\end{eqnarray}
It is worth noting that $\ket{\widetilde{\Psi}_i(t)}$ satisfies Eq.~(\ref{SE-AS}), so does any state of the form
\begin{eqnarray}\label{form}
\ket{0}\otimes\ket{\psi(t)}+ \ket{1}\otimes
  M(t)\ket{\psi(t)},
\end{eqnarray}
where $\ket{\psi(t)}$ denotes any state of $S$ satisfying Eq.~(\ref{SE}).

To proceed, we let the density operator of the Hermitian system be
\begin{eqnarray}\label{AS:density-operator}
\rho(t):=\sum_{i=1}^{n} p_{i}\ket{\Psi_i(t)}\bra{\Psi_i(t)},
\end{eqnarray}
with
\begin{eqnarray}
\ket{\Psi_i(t)}:=\ket{0}\otimes\ket{\psi_i(t)}+
\ket{1}\otimes
M(t)\ket{\psi_i(t)},
\end{eqnarray}
where $\ket{\psi_i(t)}$, $i=1,\cdots,n$, are the states appearing in Eq.~(\ref{rho-ensemble}). Noting that all the $\ket{\Psi_i(t)}$'s are of the form
(\ref{form}), one can see that $\rho(t)$ in Eq.~(\ref{AS:density-operator}) is indeed a density operator of the Hermitian system.

Suppose that the observable to be measured is
\begin{eqnarray}
A(t):=\sum_{i=1}^d a_i(t)\ket{\widetilde{a_i}(t)}\bra{\widetilde{a_i}(t)},
\end{eqnarray}
with
\begin{eqnarray}\label{ai}
\ket{\widetilde{a_i}(t)}=\ket{0}\otimes\ket{a_i(t)}+\ket{1}\otimes M(t)\ket{a_i(t)}.
\end{eqnarray}
Note that $A(t)$ is Hermitian in the conventional sense. According to the standard quantum measurement theory, the possible outcome of this measurement is $a_i(t)$, with the probability given by
\begin{eqnarray}
\bra{\widetilde{a_i}(t)}\rho(t)\ket{\widetilde{a_i}(t)}.
\end{eqnarray}
Using Eqs.~(\ref{AS:density-operator}) and (\ref{ai}), one can verify that $\bra{\widetilde{a_i}(t)}\rho(t)\ket{\widetilde{a_i}(t)}=P_i(t)$.
Now, it is clear that this measurement has the same outcomes and the probabilities as the measurement we postulated above.

%Now, performing the projective measurement $\{\ket{0}\bra{0},\ket{1}\bra{1}\}$ on $A$ and post-selecting the outcome associated with $\ket{0}\bra{0}$,
%we obtain the following state of $S$,
%\begin{eqnarray}
%\widetilde{\rho}_S(t)=\frac{\sum_{i=1}^{n}p_i\ket{\psi_i(t)}\bra{\psi_i(t)}}
%{p(t)},
%\end{eqnarray}
%with the probability
%\begin{eqnarray}
%p(t)=\sum_{i=1}^{n}p_i\inp{\psi_i(t)}{\psi_i(t)}.
%\end{eqnarray}
%On the other hand, it follows from Eq.~(\ref{pseudo-H}) that $W(t)O(t)$ is Hermitian in the conventional sense, indicating that $\tr[W(t)O(t)\widetilde{\rho}_S(t)]$ can be measured by using another standard quantum measurement. With these two measurements, we can get
%\begin{eqnarray}\label{AS:meas}
%p(t)\tr[W(t)O(t)\widetilde{\rho}_S(t)]= \tr[O(t)\rho_S(t)].
%\end{eqnarray}
%That is, $\tr[O(t)\rho_S(t)]$ can be obtained from measured quantities $\tr[W(t)O(t)\widetilde{\rho}_S(t)]$ and $p(t)$.

\section{On $\mathcal{PT}$-symmetry breaking}
\label{sec:de}

We inspect our formulation in connection with time-independent $\mathcal{PT}$QM. To do this,
consider again the time-independent $\mathcal{PT}$-symmetric Hamiltonian $\mathcal{H}$. For simplicity, we assume that $\mathcal{H}$ is diagonalizable, which is usually the case. Then the $\ket{\varphi_i}$'s and $\bra{\phi_i}$'s can be chosen as a complete set of biorthonormal eigenvectors \cite{2002Mostafazadeh2814}; that is, they obey
\begin{eqnarray}
\inp{\phi_i}{\varphi_j}=\delta_{ij}
\quad\textrm{and}\quad \sum_{i=1}^d\ket{\varphi_i}\bra{\phi_i}=I.
\end{eqnarray}
This enables us to express $\mathcal{H}$ as
\begin{eqnarray}\label{effective-H}
\mathcal{H}=\sum_{i=1}^d\omega_i\ket{\varphi_i}\bra{\phi_i}.
\end{eqnarray}
Using Eq.~(\ref{effective-H}), one can easily verify that
\begin{eqnarray}\label{dis:eta}
\eta(t):=\sum_{i=1}^de^{i\omega_it}\ket{i}\bra{\phi_i}
\end{eqnarray}
satisfies Eq.~(\ref{eta}). Here, $\{\ket{i},i=1,\cdots,d\}$ is an arbitrary orthonormal basis (w.r.t.~the usual inner product), i.e., $\inp{i}{j}=\delta_{ij}$. Inserting Eq.~(\ref{dis:eta}) into Eq.~(\ref{metric-operator}) gives
\begin{eqnarray}\label{MO-special}
W(t)=\sum_{i=1}^de^{-2\Im(\omega_i) t}\ket{\phi_i}\bra{\phi_i}.
\end{eqnarray}
Substituting Eq.~(\ref{MO-special}) into $K(t)=-\frac{1}{2}W^{-1}(t)\partial_tW(t)$ and noting that $W^{-1}(t)=\sum_{i=1}^de^{2\Im(\omega_i) t}\ket{\varphi_i}\bra{\varphi_i}$, we obtain
\begin{eqnarray}\label{K(t)-special}
K(t)=\sum_{i=1}^d \Im(\omega_i)\ket{\varphi_i}\bra{\phi_i},
\end{eqnarray}
from which we further have
\begin{eqnarray}\label{H(t)-special}
H(t)=\sum_{i=1}^d \Re(\omega_i)\ket{\varphi_i}\bra{\phi_i}.
\end{eqnarray}
Here, $K(t)$ and $H(t)$ are actually time-independent. Throughout this paper, the symbol $t$ appearing in $K(t)$, $H(t)$, and $W(t)$ is kept even in the time-independent case, simply for notational consistency.

Now, if $\mathcal{PT}$ symmetry is not spontaneously broken, i.e., all the $\omega_i$'s are real, it follows from Eqs.~(\ref{MO-special}), (\ref{K(t)-special}), and (\ref{H(t)-special}) that $W(t)$ equals to the time-independent metric operator in Eq.~(\ref{Bender-MO}), $H(t)=\mathcal{H}$, and $K(t)=0$. As an immediate consequence, our formulation reduces to time-independent $\mathcal{PT}$QM. Besides, even if $\mathcal{PT}$ symmetry is broken, a suitable metric operator still exists but at the price of being time-dependent. In this case, $H(t)$ and $K(t)$ correspond to the real and imaginary parts of the eigenvalues of $\mathcal{H}$, respectively. Interestingly, combining these two cases, we arrive at a new understanding of spontaneous $\mathcal{PT}$-symmetry breaking: In the unbroken regime, $H(t)$ and $\mathcal{H}$ coincide and the metric operator can be chosen as being time-independent, whereas in the broken regime, $H(t)$ deviates from $\mathcal{H}$ and the corresponding metric operator is time-dependent.

To demonstrate the above finding, consider then a two-level system with the following time-independent Hamiltonian
\begin{eqnarray}\label{example1-H}
\mathcal{H}=\kappa\begin{pmatrix}
                          i\alpha & -1 \\
                          -1 & -i\alpha
                        \end{pmatrix},
\end{eqnarray}
where $\kappa$ and $\alpha$ are two positive real numbers.
This model has been studied extensively in the literature \cite{2002Bender270401,2010Rueter192}. Choosing the parity operator to be the Pauli matrix $\sigma_x$
and noting that $\mathcal{T}$ performs only complex conjugation here, one can easily verify that $\mathcal{H}$ in Eq.~(\ref{example1-H}) is indeed $\mathcal{PT}$-symmetric.  The eigenvalues of $\mathcal{H}$ are $\omega_{1,2}=\pm\kappa\sqrt{1-\alpha^2}$. Hence, the $\mathcal{PT}$ phase-transition point is $\alpha=1$. In the unbroken regime, i.e., $\alpha<1$, the biorthonormal eigenvectors can be chosen as
\begin{eqnarray}
\ket{\varphi_1^{\textrm{un}}}
&=&\frac{1}{\sqrt{2\cos\theta}}
\begin{pmatrix}
e^{-\frac{i\theta}{2}} \\
e^{\frac{i\theta}{2}}
\end{pmatrix},
~~
\ket{\varphi_2^{\textrm{un}}}
=\frac{1}{\sqrt{2\cos\theta}}
\begin{pmatrix}
e^{\frac{i\theta}{2}} \\
-e^{-\frac{i\theta}{2}}
\end{pmatrix},\nonumber\\
\ket{\phi_1^{\textrm{un}}}
&=&\frac{1}{\sqrt{2\cos\theta}}
\begin{pmatrix}
e^{\frac{i\theta}{2}} \\
e^{-\frac{i\theta}{2}}
\end{pmatrix},
~~
\ket{\phi_2^{\textrm{un}}}
=\frac{1}{\sqrt{2\cos\theta}}
\begin{pmatrix}
e^{\frac{-i\theta}{2}} \\
-e^{\frac{i\theta}{2}}
\end{pmatrix}.\nonumber
\end{eqnarray}
Here $\theta:=\arcsin(\alpha)$. Direct calculations show that
\begin{eqnarray}
W(t)=\frac{1}{\cos\theta}
\begin{pmatrix}
  1 & i\sin\theta \\
  -i\sin\theta & 1
\end{pmatrix},
\end{eqnarray}
and
\begin{eqnarray}\label{Ex1:unbroken-HK}
H(t)=\mathcal{H}, \quad K(t)=0.
\end{eqnarray}
On the other hand, in the broken regime, i.e. $\alpha>1$, the biorthonormal eigenvectors can be chosen as
\begin{eqnarray}
\ket{\varphi_1^{\textrm{br}}}
&=&\frac{1}{\sqrt{2\sinh\theta}}
\begin{pmatrix}
e^{\frac{\theta}{2}} \\
ie^{-\frac{\theta}{2}}
\end{pmatrix},
~~
\ket{\varphi_2^{\textrm{br}}}
=\frac{1}{\sqrt{2\sinh\theta}}
\begin{pmatrix}
e^{-\frac{\theta}{2}} \\
ie^{\frac{\theta}{2}}
\end{pmatrix},\nonumber\\
\ket{\phi_1^{\textrm{br}}}
&=&\frac{1}{\sqrt{2\sinh\theta}}
\begin{pmatrix}
e^{\frac{\theta}{2}} \\
-ie^{-\frac{\theta}{2}}
\end{pmatrix},
~~
\ket{\phi_2^{\textrm{br}}}
=\frac{1}{\sqrt{2\sinh\theta}}
\begin{pmatrix}
-e^{\frac{-\theta}{2}} \\
ie^{\frac{\theta}{2}}
\end{pmatrix}.\nonumber
\end{eqnarray}
Here $\theta:=\textrm{arcosh}(\alpha)$. Using Eqs.~(\ref{MO-special}), (\ref{K(t)-special}), and (\ref{H(t)-special}) and noting that $\omega_{1,2}=\pm i\kappa\sinh\theta$, we obtain, after some algebra,
\begin{eqnarray}
W(t)=\frac{1}{\sinh\theta}
\begin{pmatrix}
  \cosh\left(\theta-2\kappa t\sinh\theta\right) & i\cosh\left(2\kappa t\sinh\theta\right) \\
  -i\cosh\left(2\kappa t\sinh\theta\right) & \cosh\left(\theta+2\kappa t\sinh\theta\right)
\end{pmatrix},\nonumber
\end{eqnarray}
and
\begin{eqnarray}\label{Ex1:broken-HK}
H(t)=0,\quad K(t)=-i\mathcal{H}.
\end{eqnarray}
It is interesting to observe from Eqs.~(\ref{Ex1:unbroken-HK}) and (\ref{Ex1:broken-HK}) that the effective Hamiltonian $\mathcal{H}$ plays the role of $H(t)$ in the unbroken regime, whereas it plays the role of $K(t)$ in the broken regime.

The key message from this section is the following: From the dynamical point of view, depending on whether $\mathcal{H}$ possesses a real spectrum or not, we have either a time-independent or a time-dependent metric operator $W(t)$. In both cases, an appropriate energy observable $H(t)$ with real energy eigenvalues can be found. The consequence of  spontaneous $\mathcal{PT}$-symmetry breaking is hence no longer the emergence of imaginary energy values, but the necessity of introducing a time-dependent metric operator.

\section{Applications in quantum thermodynamics}
\label{sec:im}

Let us now illustrate the possible applications of our framework with two well-known examples in quantum thermodynamics. From both examples, it can be seen that our framework not only offers an alternative interpretation of the dynamics of non-Hermitian systems, but also triggers an intriguing extension of thermodynamical relations from $\mathcal{PT}$-symmetric systems to generic non-Hermitian systems.

\subsection{Jarzynski equality}

The JE relates free energy differences between two equilibrium states to the exponentiated work averaged over an ensemble of trajectories. It is one of the celebrated results in thermodynamics \cite{2009Esposito1665,2011Campisi771} and was recently generalized to $\mathcal{PT}$-symmetric
systems \cite{2015Deffner150601,2016Gardas23408}.
Here, we further extend JE to generic non-Hermitian systems.

Given any $\mathcal{H}(t)$, we find a $W(t)$ satisfying Eq.~(\ref{metric-operator}) with Eq.~(\ref{eta}). Then we split $\mathcal{H}(t)$ into two parts, i.e., $K(t)=-\frac{1}{2}W^{-1}(t)\partial_tW(t)$ and $H(t)$ in Eq.~(\ref{H(t)}), and treat $H(t)$ as the energy observable. Note that $H(t)$ has a spectral decomposition
\begin{eqnarray}
H(t)=\sum_n\varepsilon_n(t)\Pi_n(t),
\end{eqnarray}
where $\varepsilon_n(t)\in\mathbb{R}$ denotes the energy eigenvalue (which can be measured via our measurement postulate) and $\Pi_n(t)$ is a physically Hermitian operator satisfying
\begin{eqnarray}
\Pi_m(t)\Pi_n(t)=\delta_{mn}\Pi_n(t)\quad\textrm{and}\quad\sum_n\Pi_n(t)=I.
\end{eqnarray}
In the nondegenerate case, $\Pi_n(t)=\ket{\varepsilon_n(t)}\bra{\varepsilon_n(t)}W(t)$, with $\ket{\varepsilon_n(t)}$ denoting the eigenvector associated with $\varepsilon_n(t)$.

%Following Refs.~\cite{2015Deffner150601,2016Gardas23408}, we may think of the expectation value $\tr[H(t)\rho_S(t)]$ as arising from a projective energy measurement. Specifically,  $\tr[H(t)\rho_S(t)]=\sum_n \varepsilon_n(t)\tr[\Pi_n(t)\rho_S(t)]$, leading to the interpretation that $\tr[H(t)\rho_S(t)]$ arises from the projective measurement with $\varepsilon_n(t)$'s labeling its possible outcomes and $\tr[\Pi_n(t)\rho_S(t)]$ being the probability of getting $\varepsilon_n(t)$.

Now, consider the following process \cite{2009Esposito1665,2011Campisi771}: At $t=0$, $S$ is prepared in the Gibbs state
\begin{eqnarray}\label{Im1:is}
\rho_S(0)=\frac{e^{-\beta H(0)}}{Z(0)},
\end{eqnarray}
where $Z(0):=\tr \exp[-\beta H(0)]$ is the partition function, and its energy is measured through the measurement associated with $H(0)$. Then, $S$ is subjected to the evolution governed by Eq.~(\ref{LE}), before the second measurement associated with $H(\tau)$ is performed at $t=\tau$. For this process, the probability that the two outcomes $\varepsilon_m(0)$ (at $t=0$) and $\varepsilon_n(\tau)$ (at $t=\tau$) jointly occur reads \cite{2015Deffner150601,2016Gardas23408}
\begin{eqnarray}\label{Im1:jp}
&&P[\varepsilon_m(0),\varepsilon_n(\tau)]\nonumber\\
&&=\tr\left[\Pi_n(\tau)
\mathcal{U}(\tau)\Pi_m(0)\rho_S(0)\Pi_m(0)\mathcal{U}^{-1}(\tau)\right].
\end{eqnarray}
Here, $\mathcal{U}(\tau):=T\exp[-i\int_0^\tau\mathcal{H}(t)dt]$
is the evolution operator, where $T$ denotes the time-ordering operator.

Associated with the probability in Eq.~(\ref{Im1:jp}) is a trajectory, for which the work done to the system is given by
\begin{eqnarray}\label{Im1:w}
w=\varepsilon_n(\tau)-\varepsilon_m(0).
\end{eqnarray}
Then the average exponentiated work reads
\begin{eqnarray}\label{Im1:aw}
\expt{e^{-\beta w}}:=\sum_{mn}P[\varepsilon_m(0),\varepsilon_n(\tau)]
e^{-\beta[\varepsilon_n(\tau)-\varepsilon_m(0)]}.
\end{eqnarray}
Substituting Eq.~(\ref{Im1:jp}) into Eq.~(\ref{Im1:aw}) and after some algebra (see Appendix \ref{app:A}), we arrive at the JE:
\begin{eqnarray}\label{Im1:JE}
\expt{e^{-\beta w}}=e^{-\beta\Delta F}.
\end{eqnarray}
Here, $\Delta F:=F(\tau)-F(0)$, with
\begin{eqnarray}
F(0)=-\frac{1}{\beta}\ln Z(0)\quad\textrm{and}\quad F(\tau)=-\frac{1}{\beta}\ln Z(\tau)
\end{eqnarray}
denoting respectively the free energies for the states $\exp[-\beta H(0)]/Z(0)$ and $\exp[-\beta H(\tau)]/Z(\tau)$,
where $Z(\tau):=\tr \exp[-\beta H(\tau)]$.

It is worth noting that our finding is more general than what was obtained in the previous studies \cite{2015Deffner150601,2016Gardas23408}.
Indeed, our finding shows that JE holds for generic (finite-dimensional) non-Hermitian systems, whereas those previous studies \cite{2015Deffner150601,2016Gardas23408} had concluded that JE does not hold for $\mathcal{PT}$-symmetric systems in the broken phase.

\subsection{Crooks fluctuation theorem}

The CFT belongs to the earliest discovered class of fluctuation theorems in thermodynamics \cite{2009Esposito1665,2011Campisi771}. Recently, it was generalized to $\mathcal{PT}$-symmetric
systems \cite{2017Zeng31001}. Here, using our framework as well as following Ref.~\cite{2017Zeng31001}, we further extend CFT to generic non-Hermitian systems.

In deference to common usage, we refer to the process mentioned in the above paragraph as the forward process. For this process, the probability distribution of work values is given by
\begin{eqnarray}\label{Im2:p(w)}
p(w)=\sum_{mn}\delta[w-\varepsilon_n(\tau)+\varepsilon_m(0)]
P[\varepsilon_m(0),\varepsilon_n(\tau)],
\end{eqnarray}
where $\delta(x)$ is the Dirac $\delta$ function. Conjugate to this forward process is the so-called time-reversed process, which can be described as follows:
Initially, $S$ is prepared in the state
\begin{eqnarray}\label{Im2:is}
\rho_S^{\textrm{tr}}(0)=\frac{e^{-\beta H(\tau)}}{Z(\tau)},
\end{eqnarray}
and its energy is measured through the measurement associated with  $H(\tau)$. Next, $S$ undergoes the time-reversed evolution until $t=\tau$, i.e., $\rho_S^{\textrm{tr}}(\tau)=\mathcal{U}^{-1}(\tau)\rho_S^{\textrm{tr}}(0)
\mathcal{U}(\tau)$ \cite{2009Esposito1665,2011Campisi771}. Lastly, the second measurement associated with $H(0)$ is performed at $t=\tau$.

For the time-reversed process, the probability that the two outcomes $\varepsilon_n(\tau)$ (at $t=0$) and $\varepsilon_m(0)$ (at $t=\tau$) jointly occur is given by
\begin{eqnarray}\label{Im2:tr-jp}
&&P^\textrm{tr}[\varepsilon_n(\tau),\varepsilon_m(0)]\nonumber\\
&&=\tr\left[\Pi_m(0)
\mathcal{U}^{-1}(\tau)\Pi_n(\tau)\rho_S^{\textrm{tr}}(0)
\Pi_n(\tau)\mathcal{U}(\tau)\right].
\end{eqnarray}
Corresponding to this probability, the work done to the system is $\varepsilon_m(0)-\varepsilon_n(\tau)$.
Then the probability distribution of work values reads
\begin{eqnarray}\label{Im2:tr-p(w)}
p^\textrm{tr}(w)=\sum_{mn}\delta[w-\varepsilon_m(0)+\varepsilon_n(\tau)]
P^\textrm{tr}[\varepsilon_n(\tau),\varepsilon_m(0)].\nonumber\\
\end{eqnarray}
Direct calculations (see Appendix \ref{app:A}) show that
\begin{eqnarray}\label{Im2:eq1}
P[\varepsilon_m(0),\varepsilon_n(\tau)]=
P^\textrm{tr}[\varepsilon_n(\tau),\varepsilon_m(0)]
e^{\beta(w-\Delta F)}.
\end{eqnarray}
Here $w$ is given by Eq.~(\ref{Im1:w}).
Using Eqs.~(\ref{Im2:p(w)}), (\ref{Im2:tr-p(w)}), and (\ref{Im2:eq1}), we reach the CFT:
\begin{eqnarray}
\frac{p(w)}{p^\textrm{tr}(-w)}=e^{\beta(w-\Delta F)}.
\end{eqnarray}

Interestingly, although the previous study \cite{2017Zeng31001} had concluded that the CFT breaks down for $\mathcal{PT}$-symmetric systems in the broken phase, our formulation is able to rescue the CFT for arbitrary (finite-dimensional) non-Hermitian systems, thanks to a suitable metric
operator found here.
As the final remark of this section, applications of our formulation in quantum thermodynamics are not limited to the specific examples discussed here. For example, the universal quantum work relation \cite{2008Andrieux230404,2018Wei12114} can also be extended under our framework.

\section{Concluding remarks}
\label{sec:con}

It should be fruitful for future work to find  more applications of our formulation. As one interesting example, let us consider the question about the relative phase between two states $\ket{\psi(0)}$ and $\ket{\psi(\tau)}$ at two different times $t=0$ and $t=\tau$. This turns out to be highly nontrivial, as the Hilbert space in our formulation is always tagged with the time parameter $t$, because of the time-dependence of its inner product. Only when $W(\tau)=W(0)$, we may view states $|\psi(0)\rangle$ and $|\psi(\tau)\rangle$ as two vectors living in the same Hilbert space and as such their relative phase can be evaluated without ambiguity.  In the special case of
$\ket{\psi(\tau)}=e^{i\alpha}\ket{\psi(0)}$, we may further divide the phase angle $\alpha$ into a dynamical phase and the Aharonov-Anandan (AA) phase \cite{1987Aharonov1593}.  Specifically, we may extend the geometric phase formalism developed for $\mathcal{PT}$-symmetric systems \cite{2018Zhang,2018Zhang-accompanying} to arbitrary (finite-dimensional) non-Hermitian systems under the condition of $W(\tau)=W(0)$. To do this, define an auxiliary state
\begin{eqnarray}
\ket{\phi_a(t)}:=e^{-if(t)}\ket{\psi(t)},
\end{eqnarray}
with $f(\tau)-f(0)=\alpha$. By definition, $\ket{\phi_a(\tau)}=\ket{\phi_a(0)}$; that is, $\ket{\phi_a(t)}$ is a cyclic gauge.
Substituting $\ket{\psi(t)}=e^{if(t)}\ket{\phi_a(t)}$ into Eq.~(\ref{SE}), splitting $\mathcal{H}(t)$ into $H(t)$ and $K(t)$, and directly using the same arguments as in Ref.~\cite{2018Zhang-accompanying} (see also Appendix \ref{app:A}), we have
\begin{eqnarray}\label{Im3:tp}
\alpha=\beta+\gamma,
\end{eqnarray}
with
\begin{eqnarray}
\beta &:=& -\int_0^\tau\ud t\inpp{\phi_a(t)}{H(t)
\phi_a(t)}_t,\label{eq:DP}\\
\gamma &:=& -\Im\int_0^\tau\ud t
\inpp{\phi_a(t)}{\dot{\phi}_a(t)}_t.\label{eq:GP}
\end{eqnarray}
Since $H(t)$ is the energy observable by construction, $\beta$ in Eq.~(\ref{eq:DP}) apparently represents a dynamical phase. On the contrary, the phase $\gamma$, as a factor obtained by removing the dynamical phase from the total phase, is of geometric nature. Indeed, $\gamma$ in Eq.~(\ref{eq:GP}) is essentially of the same form as the AA phase \cite{1987Aharonov1593} (with the only difference being a modified inner product), and moreover, it admits the geometric interpretation detailed in Ref.~\cite{2018Zhang-accompanying}. Besides,
it is also worth noting that $\gamma$ is always real and different from the complex geometric phase
defined previously \cite{1988Garrison177}.

In conclusion, we have extended $\mathcal{PT}$QM to generic (finite-dimensional) non-Hermitian systems. Our idea is to introduce a suitable time-dependent metric operator to redefine the inner product, based on which the unitarity of time evolution, the reality of energy eigenvalues, and the quantum measurement postulate are all restored. This provides a theoretical framework for understanding non-Hermitian systems from a quantum-mechanical point of view. It gives a new understanding of spontaneous $\mathcal{PT}$-symmetry breaking, that is, the consequence of spontaneous $\mathcal{PT}$-symmetry breaking can be alternatively viewed as the necessity of introducing a time-dependent metric operator. In demonstrating the applications of our formulation, we have extended two well-known results on the JE and the CFT to generic non-Hermitian systems, with interesting physical insights that previous studies were unaware of.  It should be stressed that the applications presented in this work may be experimentally observed by dilating a non-Hermitian system into a larger Hermitian system, as described in Sec.~\ref{sec:tf}.

\begin{acknowledgments}
J.G.~is supported by the
Singapore NRF Grant No.~NRF-NRFI2017-04 (WBS No.~R-144-000-378-281).
Q.W.~is supported by Singapore Ministry of Education Academic
Research Fund Tier I (WBS No.~R-144-000-352-112).
D.-J.Z.~acknowledges support from the National Natural Science Foundation of
China through Grant No.~11705105 before he joined NUS.
\end{acknowledgments}

\appendix
\setcounter{equation}{0}

\section{Proofs of Eqs.~(\ref{Im1:JE}), (\ref{Im2:eq1}), and (\ref{Im3:tp})}\label{app:A}

\textit{Proof of Eq.~(\ref{Im1:JE}):} Substituting Eq.~(\ref{Im1:is}) into Eq.~(\ref{Im1:jp}) and noting that $\exp[-\beta H(0)]=\sum_n\exp[-\beta\varepsilon_n(0)]\Pi_n(0)$, we have
\begin{eqnarray}\label{App:eq1}
P[\varepsilon_m(0),\varepsilon_n(\tau)]
=
\frac{e^{-\beta\varepsilon_m(0)}}{Z(0)}\tr\left[\Pi_n(\tau)
\mathcal{U}(\tau)\Pi_m(0)\mathcal{U}^{-1}(\tau)\right].\nonumber\\
\end{eqnarray}
Inserting Eq.~(\ref{App:eq1}) into Eq.~(\ref{Im1:aw}) gives
\begin{eqnarray}\label{App:eq2}
\expt{e^{-\beta w}}&=&\sum_{mn}
\frac{e^{-\beta\varepsilon_n(\tau)}}{Z(0)}\tr\left[\Pi_n(\tau)
\mathcal{U}(\tau)\Pi_m(0)\mathcal{U}^{-1}(\tau)\right]\nonumber\\
&=&\frac{1}{Z(0)}\sum_ne^{-\beta\varepsilon_n(\tau)}\tr[\Pi_n(\tau)]
\nonumber\\
&=&\frac{Z(\tau)}{Z(0)},
\end{eqnarray}
where the equalities $\sum_m\Pi_m(0)=I$ and $Z(\tau)=\tr\exp[-\beta H(\tau)]$ has been used.
Noting that $Z(0)=\exp[-\beta F(0)]$ and $Z(\tau)=\exp[-\beta F(\tau)]$, we can rewrite the RHS of
Eq.~(\ref{App:eq2}) as $\exp\{-\beta[F(\tau)-F(0)]\}$, thus proving Eq.~(\ref{Im1:JE}).

\textit{Proof of Eq.~(\ref{Im2:eq1}):}
Substituting Eq.~(\ref{Im2:is}) into Eq.~(\ref{Im2:tr-jp}) and noting that $\exp[-\beta H(\tau)]=\sum_n\exp[-\beta\varepsilon_n(\tau)]\Pi_n(\tau)$, we have
\begin{eqnarray}\label{App:eq3}
P^\textrm{tr}[\varepsilon_n(\tau),\varepsilon_m(0)]
=
\frac{e^{-\beta\varepsilon_n(\tau)}}{Z(\tau)}\tr\left[\Pi_m(0)
\mathcal{U}^{-1}(\tau)\Pi_n(\tau)\mathcal{U}(\tau)\right].\nonumber\\
\end{eqnarray}
Using the cyclic property of the trace, we can rewrite Eq.~(\ref{App:eq3}) as
\begin{eqnarray}\label{App:eq4}
P^\textrm{tr}[\varepsilon_n(\tau),\varepsilon_m(0)]
=
\frac{e^{-\beta\varepsilon_n(\tau)}}{Z(\tau)}\tr
\left[\Pi_n(\tau)\mathcal{U}(\tau)\Pi_m(0)
\mathcal{U}^{-1}(\tau)\right].\nonumber\\
\end{eqnarray}
From Eqs.~(\ref{App:eq1}) and (\ref{App:eq4}), it follows immediately that
\begin{eqnarray}\label{App:eq5}
P[\varepsilon_m(0),\varepsilon_n(\tau)]=
P^\textrm{tr}[\varepsilon_n(\tau),\varepsilon_m(0)]
e^{\beta w}\frac{Z(\tau)}{Z(0)}.
\end{eqnarray}
Rewriting Eq.~(\ref{App:eq5}) using $Z(0)=\exp[-\beta F(0)]$ and $Z(\tau)=\exp[-\beta F(\tau)]$, we obtain Eq.~(\ref{Im2:eq1}). This completes the proof.

\textit{Proof of Eq.~(\ref{Im3:tp}):}
Substituting $\ket{\psi(t)}=e^{if(t)}\ket{\phi_a(t)}$ into Eq.~(\ref{SE}) [with $H(t)$ and $K(t)$ identified from $\mathcal{H}(t)$] and contracting its both sides with $\bra{\phi_a(t)}W(t)$, we have
\begin{eqnarray}
&&\dot{f}(t)=\nonumber\\
&&-\inpp{\phi_a(t)}{[H(t)+iK(t)]
\phi_a(t)}_t
+i\inpp{\phi_a(t)}{\dot{\phi}_a(t)}_t,\nonumber
\end{eqnarray}
where the dot denotes the time derivative. Noting that $\inpp{\phi_a(t)}{K(t)
\phi_a(t)}_t=[\inpp{\dot{\phi}_a(t)}
{\phi_a(t)}_t+\inpp{\phi_a(t)}
{\dot{\phi}_a(t)}_t]/2$, we can rewrite the above equation as
\begin{eqnarray}
\dot{f}(t)=
-\inpp{\phi_a(t)}{H(t)\phi_a(t)}_t
-\Im
\inpp{\phi_a(t)}{\dot{\phi}_a(t)}_t.\nonumber
\end{eqnarray}
Integrating this equation gives Eq.~(\ref{Im3:tp}) with Eqs.~(\ref{eq:DP}) and (\ref{eq:GP}).

%\bibliography{refs.bib}

\begin{thebibliography}{48}
\bibitem [{1no()}]{1note}%
  \BibitemOpen
  \href@noop {} {}\bibinfo {note} {Given a $\eta(t)$ satisfying
  Eq.~(\ref{eta}), it is easy to see that $\lambda\eta(t)$, for any
  $\lambda\in\mathbb{C}$, satisfies Eq.~(\ref{eta}) as well. So,
  $W(t)=\abs{\lambda}^2\eta^\dagger(t)\eta(t)$, and the condition $W(t)\geq I$
  can be fulfilled as long as $\abs{\lambda}$ is large enough.}\BibitemShut
  {Stop}%
\end{thebibliography}

\end{document}